# Zero-Energy-Device for 6G: First Real-Time Backscatter Communication thanks to the Detection of Pilots from an Ambient Commercial Cellular Network

Papis Ndiaye\*, Dinh-Thuy Phan-Huy\*, Ayman Hassan\*\*, Jingyi Liao†, Xiyu Wang†, Kalle Ruttik†, Riku Jantti
\*Orange Innovation, Networks, Chatillon, France
Email: {idrissa.ndiaye, dinhthuy.phanhuy}@orange.com
\*\*Benha Faulty of Engineering, Benha University
Orange International Center, Cairo, Egypt
Email: ayman.mohamed@bhit.bu.eg.edu, ayman.hassan@orange.com
†Department of Communications and Networking, Aalto University, 02150 Espoo, Finland
Email: { jingyi.liao, xiyu.wang, kalle.ruttik, riku.jantti}@aalto.fi

*Abstract*— Ambient backscatter communication technology (AmBC) and a novel device category called zero-energy devices (ZED) have recently emerged as potential components for the forthcoming 6th generation (6G) networks. A ZED communicates with a smartphone without emitting additional radio waves, by backscattering ambient waves from base stations. Thanks to its very low consumption, a ZED powers itself by harvesting ambient light energy. However, the time variations of data traffic in cellular networks prevents AmBC to work properly. Recent works have demonstrated experimentally that a backscatter device could be detected by listening only ambient pilot signals (which are steady) instead of the whole ambient signal (which is bursty) of 4G. However, these experiments were run with a 4G base station emulator and a bulky energy greedy backscatter device. In this paper, for the first time, we demonstrate real-time AmBC on the field, with Orange commercial 4G network as ambient source and Orange Zero-Energy Device.

*Keywords—Ambient backscatter, Ambient IoT, LTE, 6G, zero-energy-device, real-time demonstrator.*

## I. Introduction

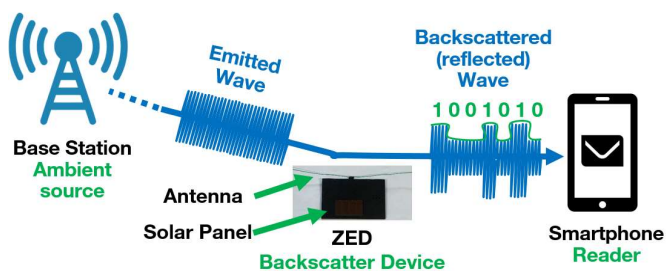

Fig. 1. AmBC and ZED principle.

Recently the 3rd Generation Partnership Project (3GPP) has started studying the new concept of Ambient Internet of Thing (IoT) [1], which aims at introducing self-powered IoT devices in 5th generation (5G) networks. Following the same trend, the European Flagship project on 6th generation (6G) Hexa-X, lists the new concept of Zero-Energy-Device (ZED) [2] among its proposed key innovations for 6G [3]. The ZED is an IoT device that harvests ambient light to power itself and that exploits the Ambient Backscatter (AmBC) technology (initially tested with ambient TV signals [4]) to communicate. As illustrated in Fig. 1, the ZED reflects (backscatters) waves of an ambient cellular network to send a message to a smartphone. The reflection is modulated over time, like in passive RFID technology. Like an RFID reader, the smartphone "reads" the ZED message in the variations in the received ambient signal. However, unlike RFID technology where dedicated portals are needed to read a backscattered message, AmBC uses wireless networks. Hence, the ZED just needs to be within network coverage to be detected by smartphones. Note that in 3GPP [1], this ZED concept is like the so-called "UE-assisted Ambient IoT" concept.

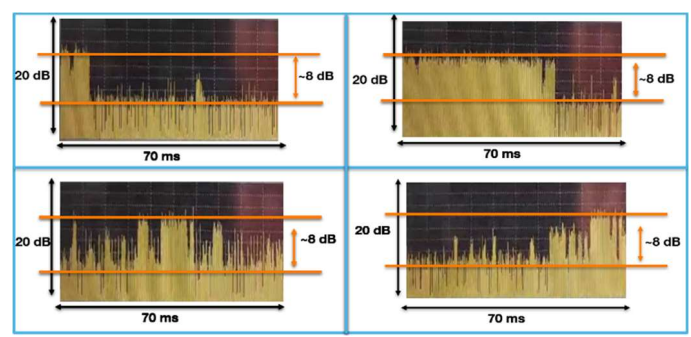

Fig. 2. Typical received power variations in time, due to data traffic burstiness, measured in the 4G 10 MHz band at 768 MHz.

Hence, AmBC technology and ZED have a large potential impact in IoT services in general [5] and logistic tracking in particular [2]. However, the ZED detection strongly relies on the steadiness of the ambient source. Unfortunately, the data traffic of cellular networks is highly bursty. Fig. 2 illustrates a typical received power over time, measured in a 4G downlink (DL) band (768 MHz), which strongly varies over time.

Recent works have experimentally demonstrated the feasibility of AmBC for 4G (as 4G is the ambient cellular network with the largest coverage available today) by listening to the long-term evolution (LTE) standardized pilots called Cell-Specific Reference Signals (CRS) (which are sent periodically, even in the absence of data traffic) to estimate the channel and demodulate the backscattered signal from the obtained channel impulse response, instead of relying on the time varying overall LTE signal [6-8]. In these works, the authors use frequency-shift keying (FSK) modulation to separate the source-to backscatter device-to receiver channel from the direct source-to-receiver channel. However, in these works, a 4G Base Station (BS) emulator is used as an ambient source, the propagation channel is controlled, and the backscatter device (BD) is bulky and not energy-autonomous.

In this paper, for the first time, we demonstrate a real-time AmBC on the field, with Orange live commercial 4G network as ambient source (instead of a 4G BS emulator) and Orange Zero-Energy Device prototype [2] (instead of a bulky non energy-autonomous BD), and an adaptation of the real-time receiver of [6-8] to reach detection on the field.

Section II presents the demonstration Setup, Section III our measurement results and Section IV concludes this paper.

## II. DEMONSTRATION SETUP DESCRIPTION

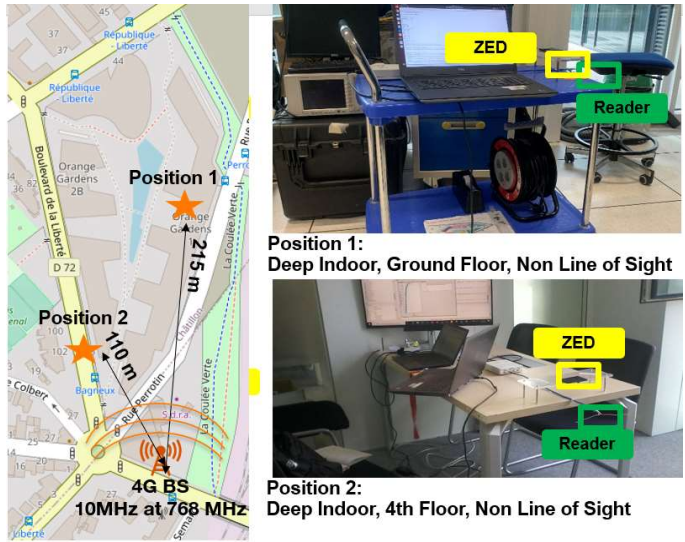

Fig. 3. Measurement setup and measurement positions.

Fig. 3 shows our experimental set-up and the positions of two field measurements. In this setup, the ambient source is a base station from Orange live commercial 4G network, with a DL 10 MHz band at 768 MHz. **The traffic of the BS and the radio propagation conditions are completely out of our control, during the experiments.**

The ZED developed in [2] is used for this experiment is re-programmed to send FSK as the BD in [6-8]. The FSK symbol duration is 40ms. The frequencies F0=125Hz and F1=500Hz are used to code for bits '0' and '1'. The ZED sends periodically, a frame of 120 bits composed of a synchronization sequence followed by a 57-bits data sequence. The resulting frame lasts 4.8 seconds.

The real-time receiver is a replication of the one presented in [6-8]. It is a laptop connected to a universal software radio peripheral (USRP B210), and it runs a slighty modifie version of the open-source code indicated by [6-8] to demodulates the ZED message in real-time. Fig. 4 illustrates the receiver signal processing blocks. As in [6-8], it estimates the channel thanks to CRS pilots, detects FSK bits, then, performs frame synchronization and then bit error rate (BER) computation. Differently from [6-9-8], we implement a different and longer synchronization sequence, to get a better detection probability on the field.

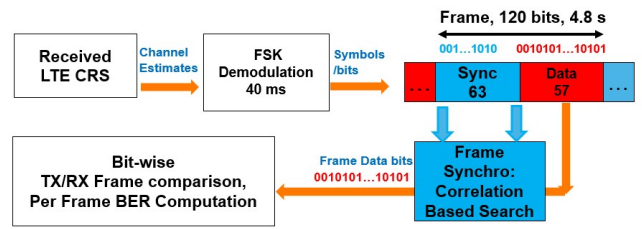

Fig. 4. Real-Time Receiver.

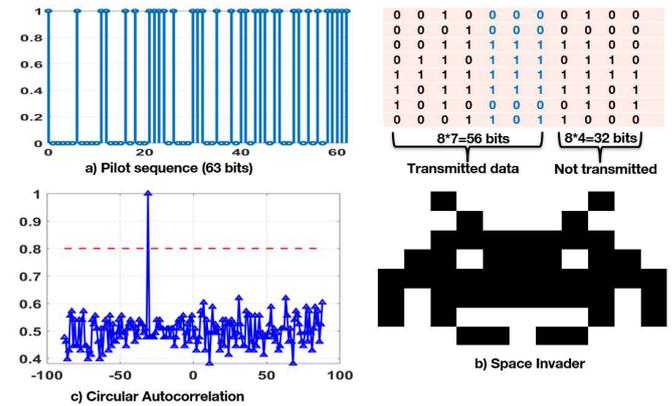

Fig. 5. Sequences: Synchronisation sequence (a), Correlation of synchronisation sequence with full sequence (sync and data) (b), data sequence (c).

Fig. 5-a) illustrates the structures of the new pilot sequence. The chosen sync sequence is a 63-bits maximal-length sequence (instead of 21-bits in [6-8]) generated using the liner-feedback shift-register (LFSR) generating polynomial $X^6+X^5+1$ [9]. As illustrated by Fig. 5-b), the data sequence codes for the partial image of Space Invader and the full Space Invader picture can be reconstituted by symmetry. Also, in Fig. 5-c), we plot circular autocorrelation between the transmitted binary sequence (composed of pilots and data) and the isolated synchronization sequence). In our frame synchronization mechanism, we set the threshold of correlation to 0.8. A frame is therefore detected as long as the received synchronization sequence is received with a level of correlation higher than 0.8, or equivalently, with 12 or less erroneous bits.

## III. EXPERIMENTAL RESULTS

Fig. 6 illustrates the detected frames (after the synchronization block of Fig. 3) during the measurements, as a function of the time, for positions 1 and 2, respectively. The two measurements lasted for around more than 1 hour. In Position 2, frames are regularly detected every 4.8 seconds, whereas in Position 1, many frames are missed.

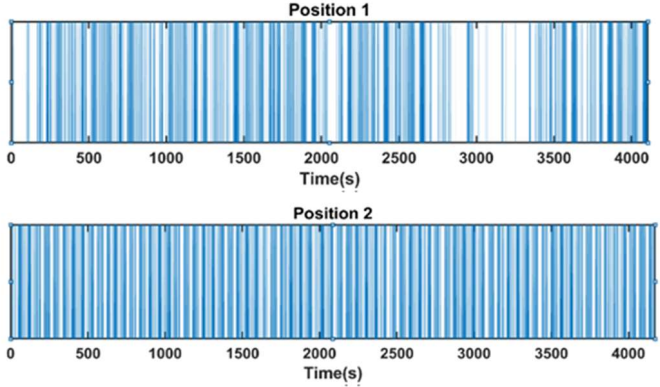

Fig. 6. Detected frames versus time.

Table I summarizes the measurement results. For each position, the ambient 4G SNR has been measured with a spectrum analyzer in max hold by scanning various positions in the room. The average SNR never exceed 0dB and 4dB in Position 1 and 2, respectively. Consequently, Position 2 has better radio conditions. As a consequence, the detected ratio (i.e. number of detected over transmitted frames) and the average data BER of detected frames is better for Position 2.

TABLE I. MEASUREMENTS RESULTS

|  | Position 1 | Position 2 |
|---|---|---|
| **Average 4G SNR** | ≤0 dB | ≤4 dB |
| **Observation duration** | 4171s | 4171s |
| **Transmitted frames** | 852 | 867 |
| **Detected frames** | 561 | 835 |
| **Detection Ratio** | 65.8% | 96.27% |
| **Average Data BER** | 0.1008 | 0.0439 |

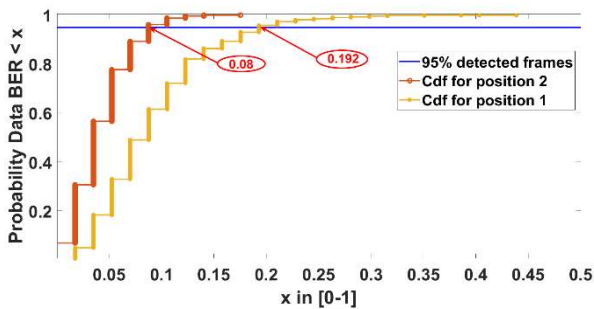

Fig. 7. Cdf of data BER of detected frames

Fig. 7 reports the cumulative density function (cdf) of the frame BER, for each position. It shows that, in good radio conditions (position 2), 95% of the detected frames have a data BER lower than 0.08, whereas in bad conditions (position 1) 95% of the detected frames have a data BER lower than 0.192. However, in both positions, 95% of the frame have a data BER lower than 0.3, which could be further corrected thanks to channel coding techniques. Also, one can notice that close to no detected frame has a data BER close to 0.5. False alarms (after frame synchronization block of Fig. 3) can therefore be assumed to be very few in these experiments.

## IV. CONCLUSION

In this paper, for the first time, we demonstrate on the field, the feasibility of Ambient Backscattering communication on cellular network, with a real-time pilot-based receiver, with a commercial 4G ambient network, and a zero-energy-device prototype. The results show that the demonstrator is robust to real network traffic and real radio propagation environment. This shows that the concept of ZED, and backscattering is promising for future cellular networks, UE-assisted Ambient IoT in 5G, and the future 6G. Future works will use such real-time demonstrator for extensive performance and coverage campaigns.

ACKNOWLEDGMENT

This work is in part supported by the European Project Hexa-X II under (grant 101095759). We thank our colleagues from Orange France for their support.